\documentclass[amsmath,amssymb,aps,preprintnumbers,showpacs,superscriptaddress,twocolumn]{revtex4}

\usepackage{graphicx}
\usepackage{dcolumn}% Align table columns on decimal point
\usepackage{bm}% bold math

\begin{document}

%\preprint{Doenitz-PRL05}

\title{Sheet-current distribution in
a dc SQUID washer probed by vortices}

\author{Dietmar Doenitz}
\author{Matthias Ruoff}
\affiliation{%
  Physikalisches Institut -- Experimentalphysik II,
  Universit\"{a}t T\"{u}bingen,
  Auf der Morgenstelle 14,
  72076 T\"{u}bingen, Germany
}
\author{Ernst Helmut Brandt}
\affiliation{%
  Max-Planck-Institut f\"{u}r Metallforschung,
  D-70506 Stuttgart,
  Germany
}
\author{John R.~Clem}
\affiliation{
 Ames Laboratory - DOE and Department of Physics and Astronomy,
 Iowa State University,
 Ames Iowa 50011, USA
}
\author{Reinhold Kleiner}
\author{Dieter Koelle}
\email{koelle@uni-tuebingen.de}
\affiliation{%
  Physikalisches Institut -- Experimentalphysik II,
  Universit\"{a}t T\"{u}bingen,
  Auf der Morgenstelle 14,
  72076 T\"{u}bingen, Germany
}

\date{\today}% It is always \today, today,

\begin{abstract}
We present a novel method, based on vortex imaging by
low-temperature scanning electron microscopy (LTSEM), to directly
image the sheet-current distribution in YBa$_2$Cu$_3$O$_7$ dc
SQUID washers. We show that the LTSEM vortex signals are simply
related to the scalar stream function describing the vortex-free
circulating sheet-current distribution $\bm J$. Unlike previous
inversion methods that infer the current distribution from the
measured magnetic field, our method uses pinned vortices as local
detectors for $\bm J$. Our experimental results are in very good
agreement with numerical calculations of $\bm J$.
\end{abstract}

% restriction on length of abstract:
% max 600 characters, including spaces
% makes roughly seven lines

%
% 68.37.Hk Scanning electron microscopy (SEM) (including EBIC)
%%%%%%%%%%%%%%%%% 74: Superconductivity %%%%%%%%%%%%%%%%%
%%%%%%%%%%%%%%%%% 74.25.-q Properties of type I and type II superconductors
% 74.25.Op Mixed states, critical fields, and surface sheaths
% 74.40.+k Fluctuations (noise, chaos, nonequilibrium superconductivity, localization, etc.)
%%%%%%%%%%%%%%%%% 74.78.-w Superconducting films and low-dimensional structures
% 74.78.Bz High-Tc films
%%%%%%%%%%%%%%%%% 85.25.-j Superconducting devices
% 85.25.Dq Superconducting quantum interference devices (SQUIDs)

\maketitle

%\section{Introduction}
%\label{Sec:Introduction}

Spatially resolved techniques can provide important insight into
current flow, arrangement of vortices, flux pinning, and noise in
superconductors and their mutual interactions. So far there has
been only one method of imaging the current distribution in
superconductors:
%, all based on the same principle which consists of two steps.
The magnetic field distribution on top of a superconducting thin
film is measured, e.g. by magneto-optics, from which the current
distribution can then be calculated by inverting the Biot-Savart
law\cite{Jooss02}.
%, Hall probe microscopy or scanning superconducting quantum interference device
%(SQUID) microscopy.

In this paper we present a novel method to directly
image the sheet-current distribution in a
YBa$_2$Cu$_3$O$_7$ thin film. We use low-temperature
scanning electron microscopy
(LTSEM)\cite{clem80,huebener84,gross94,doderer97} to
image vortices in dc SQUID
washers\cite{straub01,doenitz04}. Most techniques for
vortex imaging, such as Lorentz
microscopy\cite{Tonomura01}, scanning SQUID
microscopy\cite{Tafuri04a, Kirtley96}, scanning Hall
microscopy\cite{Grigorenko03} or
magneto-optics\cite{Goa01} rely on the detection of
the stray magnetic field produced in close proximity
to a vortex. In contrast, vortex imaging by LTSEM is
different from those techniques, as it is based on the
electron-beam-induced apparent displacement of a
vortex, pinned at position $\bm r$ in the
$(x,y)$-plane of a SQUID washer, which is detected as
a {\em change} of stray magnetic flux $\Phi(\bm r)$
coupled to the SQUID. Hence, the contrast of the LTSEM
vortex signals directly senses $\bm\nabla\Phi(\bm r)$.
Recently, Clem and Brandt\cite{clem05} have shown that
$\Phi(\bm r)$ is proportional to the scalar stream
function $G(\bm r)$ that describes the circulating
sheet-current density $\bm J(\bm r)$ flowing in the
vortex-free case at position $\bm r$ in the SQUID
washer.
In this paper we show that this relationship allows us
to use the vortices as local detectors for $\bm J(\bm
r)$: At each position a vortex has been imaged, we can
directly determine $\bm J(\bm r)$ without complicated
calculations.

%\section{Imaging of pinned vortices in a SQUID washer}
%\label{Sec:Imaging}

\begin{figure} [tb]
\center{\includegraphics[width=8.5cm]{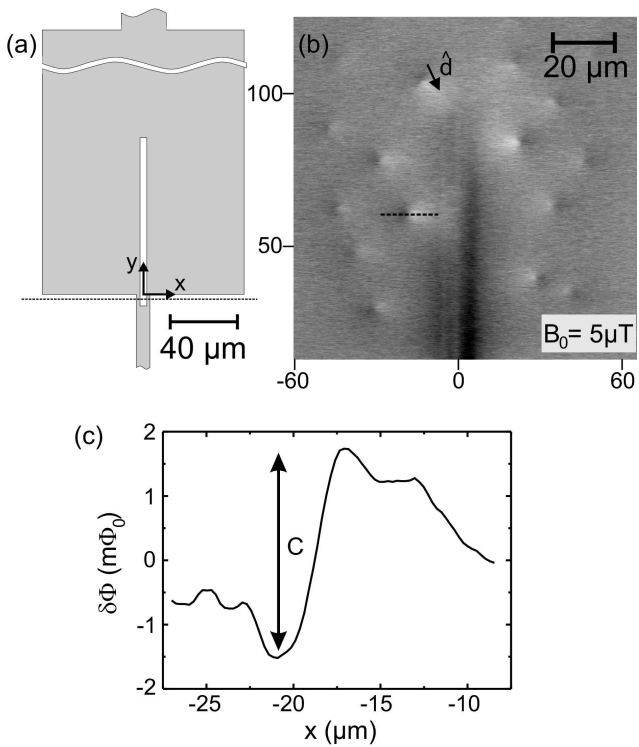}}
\caption{(a) SQUID washer design; dotted line
indicates grain boundary. (b) $\delta \Phi$ image of
washer SQUID. The arrow indicates the direction
$\hat{\bm d}$ of one of the vortex signals. Tick
labels are in units of $\mu$m. (c) Line scan $\delta
\Phi(x)$ along the dashed line shown in (b). The
signal contrast $C$, i.e., the difference between
maximum positive and negative signal from a single
vortex, is indicated by the arrow.}
\label{laybsp}
\end{figure}

In our experiments, we investigated
several dc SQUID washers [see
Fig.~\ref{laybsp}(a)] fabricated from
epitaxially grown $d$=80\,nm thick
$c$-axis oriented YBa$_2$Cu$_3$O$_7$
(YBCO) thin films. We will present an
analysis of LTSEM data obtained from one
representative device with washer size
$120\,\mu{\rm m} \times 305\,\mu{\rm m}$,
with a 100 $\mu$m long and 4 $\mu$m wide
slit. The 1 $\mu$m wide Josephson
junctions are formed by a 24$^\circ$
symmetric grain boundary in the
underlying SrTiO$_3$ substrate. For
imaging by LTSEM, the YBCO SQUIDs are
mounted on a magnetically shielded,
liquid nitrogen cooled cryostage of an
SEM \cite{gerber97a} and read out by a
standard flux-locked loop (FLL) with
3.125 kHz bias-current reversal to
eliminate $1/f$ noise due to fluctuations
in the critical current $I_c$ of the
Josephson junctions. All LTSEM images are
obtained at $T=77\,$K after cooling the
SQUID through $T_c$ in a static magnetic
field $B_0$.

%\iffalse%fig
%%%%%%%% Fig.1 %%%%%%%%%%%%%%%%%%%%%%%%%%%%%%%%%%%
%%%%%%%%%%%%%%%%%%%%%%%%%%%%%%%%%%%%%%%%%%%%%%%%
%\fi%fig

For the spatially resolved measurements, the electron beam is used
as a local perturbation that induces an increase in temperature
$\delta T(x-x_0,y-y_0)$ on the sample surface (in the $x,y$-plane)
centered on the beam spot position ($x_0,y_0)$. At $T$=77\,K the
length scale for the spatial decay of the thermal perturbation is
set by the beam-electron range $R\approx 0.5\,\mu$m for a typical
beam voltage $V_b$ = 10 kV\cite{gross94}. This gives a maximum
increase in beam-induced temperature $\Delta T$ of a few K at
$(x_0,y_0)$ for typical beam currents $I_b$ of a few nA. So-called
$\delta\Phi(x_0,y_0)$ images are obtained by recording the
e-beam-induced flux change $\delta \Phi$ in the SQUID as a
function of the e-beam coordinates $(x_0,y_0)$. To improve the
signal to noise ratio, we use a beam-blanking unit operating at
typically 5\,kHz, and the FLL output signal, i.e., the
e-beam-induced flux change in the SQUID, is lock-in detected.

The mechanism of imaging of pinned vortices can be briefly
described as follows: The e-beam-induced local increase in
temperature produces a local increase in the Pearl
length\cite{Pearl64} $\Lambda = \lambda_L^2/d$, where $\lambda_L$
is the temperature-dependent London penetration depth. Hence, the
screening currents circulating around a pinned vortex are
spatially extended due to e-beam irradiation.
If the e-beam is scanned across a vortex, the vortex's
magnetic-field distribution is distorted in the
direction of the beam spot; i.e., the center of the
distorted field distribution is displaced by $\delta
r$, which depends upon the distance of the beam spot
from the vortex.
This displacement changes the amount of stray magnetic
flux $\Phi(\bm r)$ in the negative $z$ direction that
a vortex at position $\bm r$ couples into the SQUID
hole. Hence, scanning across a vortex induces a
negative (positive) flux change $\delta \Phi$ in the
SQUID if the vortex is moved away from (towards) the
SQUID hole. Figure \ref{laybsp}(b) shows an example of
a $\delta \Phi$ image with vortices appearing as pairs
of positive (bright) and negative (dark) signals.

%\section{Relation between LTSEM vortex signal and sheet current}
%\label{Sec:Relation}

Fig.~\ref{laybsp}(c) shows a linescan $\delta \Phi(x)$
of the vortex signal along the dashed line shown in
Fig.~\ref{laybsp}(b).
The difference between the maximum positive and
negative signals from a single vortex defines the
contrast $C$ of a vortex in the $\delta \Phi$ image.
The contrast $C$ depends on the maximum vortex
displacement $\Delta r$, which is independent of the
vortex position, and on the gradient of the function
$\Phi(\bm r)$. Neglecting the (small) variation in
$|\bm\nabla \Phi(\bm r)|$ within $\Delta r$, we can
write
\begin{equation}
C = 2\Delta r \left|\bm\nabla \Phi(\bm r)\right |\; .
\label{C}
\end{equation}

We define the direction $\hat{\bm d}$ of a vortex
signal [c.f.~Fig.~\ref{laybsp}(b)] as the direction of
a unit vector in the $(x,y)$-plane pointing from the
maximum negative to maximum positive signal of the
vortex. Hence, $\hat{\bm d}$ is parallel to the
gradient of $\Phi(\bm r)$ and given as
\begin{equation}
\hat{\bm d} \equiv \bm\nabla \Phi(\bm r)/|\bm\nabla
\Phi(\bm r)|\; .
\label{d}
\end{equation}

In the vortex-free case and for currents below the
critical current of the Josephson junctions, the SQUID
washer acts as a closed superconducting loop, which
can carry a persistent supercurrent $I$, flowing
clockwise or counterclockwise. The sheet-current
density $\bm J(\bm r)$ is divergence free; thus it can
be written as
\begin{equation}
\bm J(\bm r) =I \hat{\bm z} \times
\bm\nabla G(\bm r)\; ,
\label{KG}
\end{equation}
where $\hat{\bm z}$ is the unit vector in the direction
perpendicular to the film plane and $G(\bm r)$ is a stream
function. $G(\bm r)$ depends on the Pearl length, the film
thickness and the washer geometry; it can be chosen such that
$G=0$ on the outer edge and $G=1$ on the inner edge of the
SQUID\cite{clem05}.
This situation corresponds to the ``trapped-flux
case'' (zero applied field, magnetic flux trapped in
the hole and slit, which is bridged at the edge) shown
in the upper left plot of Fig.~1 in
Ref.~\onlinecite{Brandt05a}.

While the function $\Phi(\bm r)$ applies for the
\textit{vortex state} of the SQUID washer and the
function $G(\bm r)$ applies for the
\textit{vortex-free state}, Clem and Brandt used an
energy argument to show that both functions are
closely related\citep{clem05} as
\begin{equation}
G(\bm r)=\frac{\Phi(\bm r)}{\Phi_0}\; ,
\label{clem}
\end{equation}
where $\Phi_0$=$h$/2$e$ is the magnetic flux quantum.
This equation holds as long as the current
distribution in the vicinity of a vortex is not
significantly altered by other vortices. This
requirement is well fulfilled in our experiments since
the applied fields are small (up to 40$\,\mu$T) and
thus the distance between the vortices is a few $\mu$m
or more, well above $\Lambda\approx 0.5\,\mu$m at
$T=77\,$K.

From equations (\ref{C})-(\ref{clem}) one
obtains
\begin{equation}
\bm J(\bm r) =\frac{C}{\Phi_0}\cdot\frac{I}{2\Delta
r}\cdot(\hat{\bm z}\times\hat{\bm d})\; ,
\label{zshg}
\end{equation}
i.e., $|\bm J(\bm r)| \propto C$ and $\bm J \bot \hat{\bm d}$. The
magnitude of the vortex-free sheet-current density $\bm J$ at the
position $\bm r$ of an imaged vortex is proportional to the
contrast $C$ of the LTSEM signal given by that vortex, while the
direction of $\bm J$ is perpendicular to the direction $\hat{\bm
d}$ of the vortex signal. These properties allow us to use the
pinned vortices as local probes for the magnitude and direction of
$\bm J$.

%\section{Comparison of experiment with numerical simulations}
%\label{Sec:Comparison}

\begin{figure} [b]
\vspace*{2mm}
\center{\includegraphics[width=8.5cm]{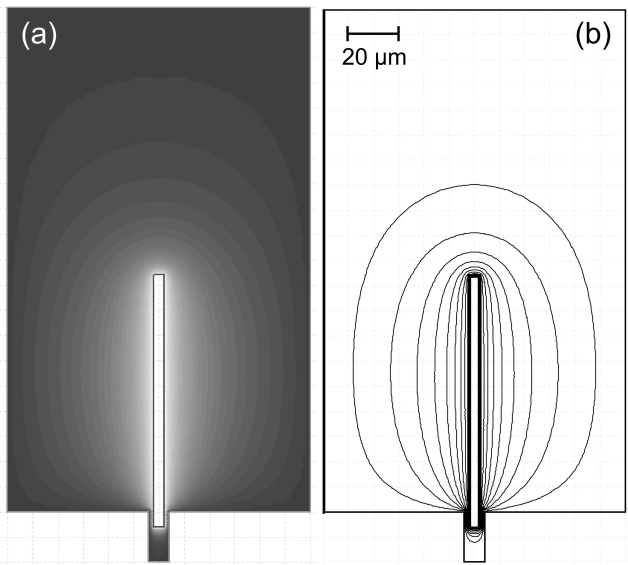}}
\caption{Numerical simulation for SQUID washer in the
vortex-free case: (a) Scalar stream function $G$;
$G=1$ (white) at the slit edge; $G=0$ (black) at the
outer edge. (b) Streamlines, showing the current
distribution in the SQUID washer.}
\label{khapaev}
\end{figure}

In order to check the validity of
(\ref{zshg}) we calculated the scalar
stream function $G(\bm r)$ and the
vortex-free sheet-current density $\bm
J(\bm r)$ for the geometry of our SQUID
washers, using the numerical simulation
software package 3D-MLSI
\citep{Khapaev03}. For the washer
geometry in Fig.~\ref{laybsp}(a), $G(\bm
r)$ and $\bm J(\bm r)$ are shown in
Fig.~\ref{khapaev}(a) and (b),
respectively.

%\iffalse%fig
%%%%%%%%% Fig.2 %%%%%%%%%%%%%%%%%%%%%%%%%%%%%%%%%%%%%%%%
%%%%%%%%%%%%%%%%%%%%%%%%%%%%%%%%%%%%%%%%%%%%%%%%%%%%%%%%%
%\fi%fig

\begin{figure*}[tbh]
\center{\includegraphics[width=16cm]{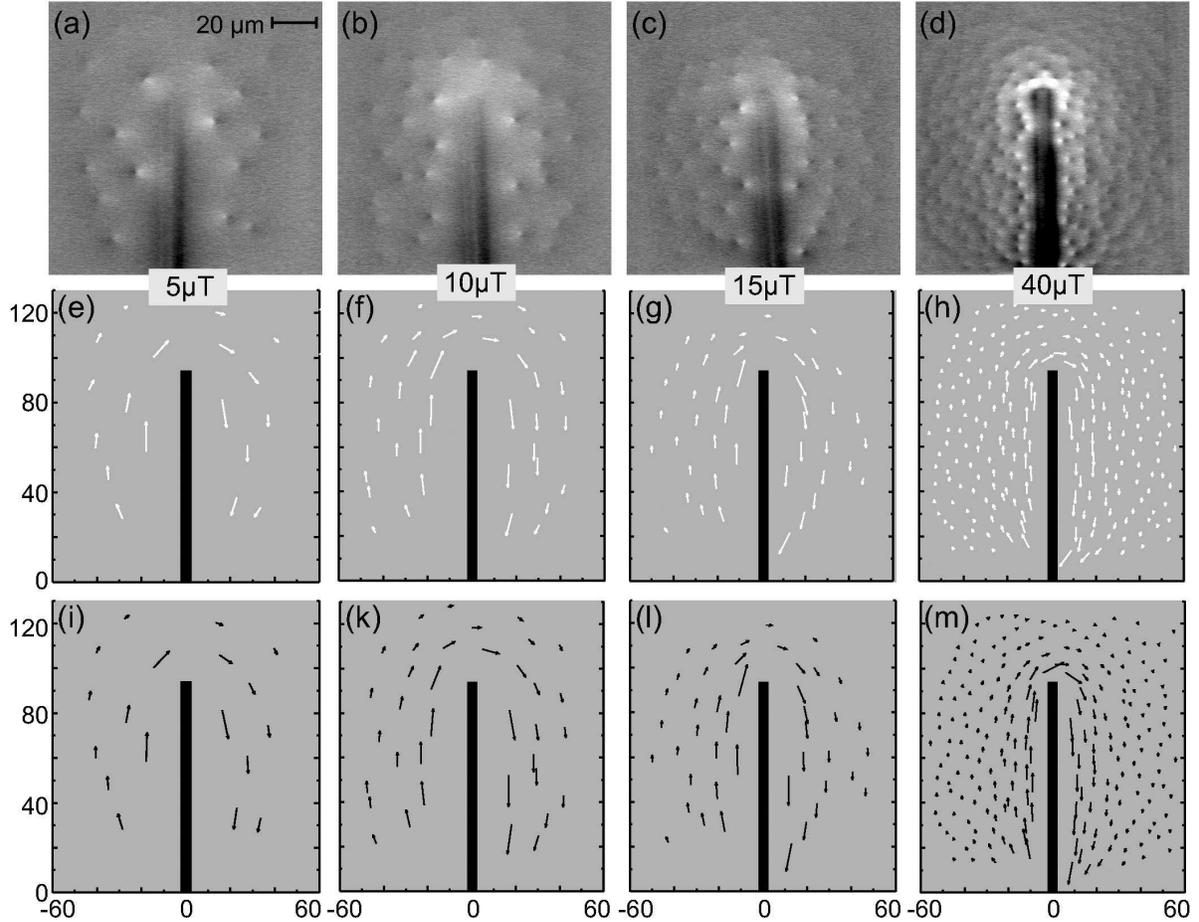}} \caption{Upper
row: LTSEM $\delta \Phi$ images for different cooling fields
$B_0$. Middle row: Magnitude and direction of sheet-current
density calculated from the corresponding $\delta \Phi$ images
(white arrows). Bottom row: Magnitude and direction of
sheet-current density calculated by numerical simulation of $\bm
J(\bm r)$ (black arrows). Tick labels are in units of $\mu$m.}
\label{4Bilder}
\end{figure*}

In Fig.~\ref{4Bilder} we compare experimental data
obtained from four $\delta \Phi$ images at different
cooling fields (upper row) with the numerically
calculated sheet-current distribution $\bm J(\bm r)$.
%\iffalse%fig
%%%%%%%%% Fig.3 %%%%%%%%%%%%%%%%%%%%%%%%%%%%%%%%%%%%%%%%
%%%%%%%%%%%%%%%%%%%%%%%%%%%%%%%%%%%%%%%%%%%%%%%%%
%\fi%fig
In Fig.~\ref{4Bilder}(e--h) (middle row) every vortex
imaged by LTSEM is represented by a white arrow at the
position of the imaged vortex. The direction of this
arrow is perpendicular to $\hat{\bm d}$, and the
length is proportional to the signal contrast $C$.
From Eq.~(\ref{zshg}) it follows that the arrows
should indicate the direction and magnitude of the
normalized vortex-free sheet-current density $\bm
J/I$. For comparison, Fig.~\ref{4Bilder}(i--m) (bottom
row) shows the results from numerical simulations;
i.e., the black arrows represent the direction and
magnitude of $\bm J(\bm r)/I$.
As shown in Fig.~\ref{4Bilder}, the agreement between
experimental data (middle row) and numerical
simulation (bottom row) is very good. The rms
deviations in magnitude and direction of the
sheet-current density are given in Table \ref{tab1}
for four values of the cooling field. Since the
proportionality constant between $C$ and $\bm J/I$
contains $\Delta r$, which depends on experimental
parameters such as e-beam power and sample
temperature, we use $\Delta r$ as a fitting parameter
for each $\delta \Phi$ image [c.f.~Table \ref{tab1}]
to obtain the best agreement between experimental and
numerical simulation data. There seems to be a trend
towards smaller values of $\Delta r$ with increasing
$B_0$. The origin of this has not been clarified yet.

%\iffalse%tab
%%%%%%%%%%% Table 1 %%%%%%%%%%%%%%%%%%%%%%%%%%%%%%
\begin{table}[t]
\caption{Rms deviation between experimentally
determined and numerically calculated sheet-current
density $\bm J(\bm r)$ and values for maximum
beam-induced displacement $\Delta r$.}
\begin{tabular}{cccc}\hline\hline
cooling field   & rms deviation & rms deviation & $\Delta r$\\
($\mu$T)        & of magnitude  & of direction  & (nm)\\ \hline
5               & 16\%          & 8.4$^\circ$   & 103\\
10              & 20\%          & 6.4$^\circ$   & 85\\
15              & 19\%          & 10.7$^\circ$  & 82\\
40              & 26\%          & 7.6$^\circ$   & 79\\
\hline\hline
\end{tabular}
\label{tab1}
\end{table}
%%%%%%%%%%%%%%%%%%%%%%%%%%%%%%%%%%%%%%%%%%%%%%%%%%
%\fi%tab
%
The deviations between experimental and simulation
data for $\bm J/I$ can easily be explained by the
difficulty to extract the direction and magnitude of
the vortex signals exactly.
Furthermore, vortices imaged close to the Josephson
junctions yield a somewhat larger deviation in the
direction of the sheet-current density as obtained
from numerical simulations. During LTSEM imaging,
shifts of the sample position up to a couple of $\mu$m
in both the $x$- and $y$-directions may occur. While
such a shift along the $x$-direction can easily be
corrected using the left and right (vertical) edges of
the SQUID washer as a reference, a shift along the
$y$-direction cannot be corrected by using the bottom
washer edge as a reference, since we avoided directly
hitting the Josephson junctions with the e-beam.
In particular, for the largest cooling field
$B_0=40\,\mu$T, the analysis of the LTSEM data yields
a significantly larger horizontal component for $|J|$
as compared to the numerically calculated current
distribution [c.f.~Fig.~\ref{4Bilder}(f)]. Most
likely, this deviation is due to a vertical shift
which occurs during LTSEM imaging.
Except for the signals close to the Josephson junctions we did not
find a systematic deviation between experimental and numerical
data.

%summary:
In conclusion, we have shown that the relation between the
vortex-generated flux $\Phi$ in a SQUID washer and the scalar
stream function $G$ allows us to use the vortices as local
detectors for the sheet-current distribution $\bm J$ in the
vortex-free case. This is a novel and more direct method of
imaging circulating supercurrents that avoids complicated
calculations. The experimental data obtained in magnetic fields up
to 40 $\mu$T are in excellent agreement with numerical
calculations of $\bm J$, confirming both the validity of our model
describing the generation of the LTSEM vortex signals and the
validity of the relationship derived by Clem and Brandt
\cite{clem05}, even in the presence of many (up to 200) vortices
in the SQUID washer.

%\begin{acknowledgments}
We thank K.~Barthel for fabricating the YBCO SQUIDs,
John Clarke for providing detailed layouts of the
Berkeley SQUID readout electronics and R.~Straub for
his support on LTSEM imaging. This work was supported
by the ESF program VORTEX, the
Landesforschungsschwerpunktsprogramm
Baden-W\"{u}rttemberg, Iowa State University of Science
and Technology under Contract No. W-7405-ENG-82 with
the U.S. Department of Energy, and the German Israeli
Research Grant Agreement (GIF) No. G-705-50.14/01.
D.~D.~ gratefully acknowledges support from the
Evangelisches Studienwerk e.V.~Villigst.
%\end{acknowledgments}

%\bibliographystyle{apsprl}
\bibliography{Stromverteilung}
\end{document}